# Organic nanofibers embedding stimuli-responsive threaded molecular components

Vito Fasano, [†] Massimo Baroncini, [||] Maria Moffa, [#] Donata Iandolo, [#,§] Andrea Camposeo, [#,*]

Alberto Credi [||,*] and Dario Pisignano[†,#,*]

[†] Dipartimento di Matematica e Fisica "Ennio De Giorgi", Università del Salento, via Arnesano
I-73100 Lecce (Italy)

[#] National Nanotechnology Laboratory of Istituto Nanoscienze-CNR, via Arnesano, I-73100
Lecce (Italy)

[||] Photochemical Nanosciences Laboratory, Dipartimento di Chimica "G. Ciamician", Università
di Bologna, via Selmi 2, I-40126 Bologna (Italy)

ABSTRACT. While most of the studies on molecular machines have been performed in solution, interfacing these supramolecular systems with solid-state nanostructures and materials is very important in view of their utilization in sensing components working by chemical and photonic actuation. Host polymeric materials, and particularly polymer nanofibers, enable the manipulation of the functional molecules constituting molecular machines, and provide a way to induce and control the supramolecular organization. Here, we present electrospun nanocomposites embedding a self-assembling rotaxane-type system that is responsive to both optical (UV-visible light) and chemical (acid/base) stimuli. The system includes a molecular axle comprised of a dibenzylammonium recognition site and two azobenzene end groups, and a dibenzo[24]crown-8 molecular ring. The dethreading and rethreading of the molecular components in nanofibers induced by exposure to base and acid vapors, as well as the photoisomerization of the azobenzene end groups, occur in a similar manner to what observed in solution. Importantly, however, the nanoscale mechanical function following external chemical stimuli induces a measurable variation of the macroscopic mechanical properties of nanofibers aligned in arrays, whose Young's modulus is significantly enhanced upon dethreading of the axles from the rings. These composite nanosystems show therefore great potential for application in chemical sensors, photonic actuators and environmentally responsive materials.

## 1. INTRODUCTION

Molecular machines are multicomponent (supramolecular) assemblies able to perform a mechanical function following an external chemical, optical, magnetic or electrical stimulus. Applications include their utilization as tweezers, catalysts, motors, switches, memories, logic gates, and valves of interest in materials science, information technology and medicine.[1-6] Threaded and interlocked molecular structures such as rotaxanes, catenanes, and related systems are most appealing platforms for the development of artificial nanoscale machines and motors.

To date, most of the studies on molecular machines have been performed in the liquid phase. Working in solution is convenient from the experimental viewpoint, allowing the molecular degrees of freedom at the base of the machine operations to be fully exploited. However, the random and variable distribution of the molecules prevents the addressing of individual molecules or given ensembles in a coherent way, which is an essential requirement to turn molecular functions into mesoscopic and macroscopic operations.[6,7] Such functions would be enabled by the interaction of the molecular machines with the environment, triggered by external stimulation of individual molecules or of a set of them, and by suitable effects on the physico-chemical properties that allow the actual molecular status to be sensed/read.[8] Therefore, there is a growing need to interface molecular machines with solid supports, by integrating them with nanostructured materials and devices, and to study the properties of these hybrid materials.

In recent years, pioneering efforts have been devoted to the incorporation of molecular machines and motors in heterogeneous matrices, and to the study of efficient strategies for delivering proper stimuli and detecting the resulting molecular changes. Examples include molecular machines deposited on surfaces as mono- or multi-layers,[9-12] embedded in solid crystals,[13] liquid crystals,[14] and gels,[15] coupled with nanoparticles,[16,17] integrated in electronic circuits,[18] and incorporated in metal-organic frameworks.[19-22] Only a few studies have been focused on polymeric host materials.[23-26] In this kind of systems, polymer molding could enable the manipulation of the functional molecules, and provide a way to induce supramolecular organization or rearrangements of the incorporated molecular guests. Moreover, polymers can convey the proper stimuli to the embedded nanomachines, and they can be used for sensing the molecular state in real time by optical, electrical or chemical signals. The large variety of polymers and nanocomposites with tailored physico-chemical properties, together with the easy processability and low cost, are further elements of interest towards this approach. For instance, azobenzene-based photochromic systems have been integrated in brush-type polymers[27] and liquid-crystalline polymer springs,[28] an approach allowing reversible, macroscopic morphological changes of the composite material to be achieved by photoisomerization of the azobenzene units.

Furthermore, the features of polymer structures embedding molecular machines can be improved via nanostructuring, in order to enhance the surface-to-volume ratio, and make the interactions with the surrounding environment more efficient.[26,29-31] In particular, polymer nanofibers show several advantages, such as their flexibility and the possibility to be easily integrated in optical or electronic devices. Moreover, the composition and structure of nanofibers made by electrospinning[32-34] can be easily tailored to target specific applied stimuli (physical and/or chemical). The high stretching rate of electrified jets favors the orientation of macromolecules, and of embedded dopants and nanoparticles as well.[35-40]

Here, we present electrospun nanocomposites embedding a self-assembling rotaxane-type system that is responsive to both optical (UV-visible light) and chemical (acid/base) stimuli.[41] The system is composed of a polyether molecular ring and a molecular axle containing a recognition site for the ring and terminated with photoactive azobenzene units. Solution studies showed that the *E-Z* photoisomerization of the terminal azobenzene units converts the system between thermodynamically stable (pseudorotaxane) and kinetically inert (rotaxane) forms.[41] Here, the reversible photoisomerization of the azobenzene units in the nanofibers is investigated by absorption spectroscopy. Moreover, the reversible dethreading and rethreading of the axle and the ring in the electrospun fibers by chemical stimulation is also assessed, using photoluminescence (PL) spectroscopy. Interestingly, we find that the macroscopic mechanical properties of the nanofiber mats embedding the pseudorotaxane complexes are influenced by their dethreading and rethreading processes at the molecular scale.

## 2. RESULTS AND DISCUSSION

**Nanofibers embedding pseudorotaxane molecular complexes**. The system is composed of a molecular axle, *EE*-**1**H•PF$_6$, comprising a dibenzylammonium recognition site and two azobenzene end groups, and a dibenzo[24]crown-8 molecular ring **2** (Fig. 1a).[41,42] Poly(methyl methacrylate) (PMMA) is used as matrix due to its good plastic behavior, processability and optical transparency (up to 93% in the visible spectral range), making it suitable for embedding photoactive compounds.[25,43,44] In organic solution, *EE*-**1**H•PF$_6$ and **2** self-assemble efficiently and rapidly to yield the pseudorotaxane [*EE*-**1**H⊂**2**]•PF$_6$.

Our nanofibers, which in their pristine form embed the [*EE*-**1**H⊂**2**]•PF$_6$ complex, can go through various photochemical or chemical processes as schematized in Fig. 1b. In order to induce the *E*→*Z* photoisomerization of the azobenzene end groups [processes (i)→(iii) and (ii) →(iv) in Fig. 1b], the nanofibers are exposed to UV light for increasing time intervals. The photoisomerization of the azobenzene end groups is reversible, and the initial state can be restored upon $Z \rightarrow E$ conversion triggered by irradiation with blue light [processes (iii)→(i) and (iv)→(ii)]. Furthermore, on the basis of solution behavior, deprotonation of the ammonium recognition site with a base is expected to cause the disassembly of the complex, as schematized in processes (i)→(ii) and (iii)→(iv) in Fig. 1b.[41,42] Eventually, treatment with an acid should cause reprotonation of the amine center and trigger rethreading of the ring and axle components [processes (ii)→(i) and (iv)→(iii)]. The dethreading and rethreading of the supramolecular complexes in the fibers, induced by exposure to base and acid vapors, respectively, are here explored with the aim of evaluating the potential of such composite nanosystems as chemical sensors and environmentally responsive materials.

Fig.s 2a and 2b show randomly oriented and nearly aligned fibers containing [$EE$-**1**H⊂**2**]•PF$_6$, that feature an average diameter of about 600 nm and 350 nm, respectively. Aligned fibers are relatively thinner and exhibit a narrower dispersion in diameter, because of the additional mechanical stretching exerted by the rotating collector on the polymeric jet.[45] The elemental analysis performed by energy dispersive X-ray spectroscopy (EDS) in different areas of the composite nanofibers (Fig. 2c) indicates a uniform distribution of dopants throughout the sample mats.

**Photoisomerization of** [$EE$-**1**H⊂**2**]•**PF$_6$-based fibers**. The occurrence of $E \rightarrow Z$ photoisomerization of the complexes induced by irradiation with UV laser light [process (i)→(iii) in Fig. 1b] is assessed by the change in the absorption spectrum of randomly-oriented fibers (Fig. 3a). The intensity of the peak at about 325 nm, characteristic of the $EE$-form of the axle component,[41] decreases with increasing UV exposure time, as shown in Fig. 3b. No more changes in the absorption intensity are observed for exposures longer than about 20 minutes, indicating that a photostationary state is reached, as commonly observed in azobenzene photoisomerization processes. On the basis of the solution absorption spectra of *EE*-**1**H•PF$_6$ and *ZZ*-**1**H•PF$_6$ it can be assumed that the absorbance of the *Z*-azobenzene unit at 325 nm is much smaller than that of the *E*-isomer and can therefore be neglected. Hence the absorption changes show that about 40% of the *E*-azobenzene units present in nanofibers are transformed to the *Z*-form. Since the two azobenzene units of *EE*-**1**H•PF$_6$ are equivalent and independent from one another, the photoreacted units are statistically distributed among the axle molecules and thus the photostationary state should contain about 44% *EE*-, 32% *EZ*-, and 24% *ZZ*-form of the axle. Exposure of the UV irradiated samples to a blue laser beam causes opposite spectral changes (Fig. 3c), indicating that the *Z*-azobenzene units are converted to the *E*-isomer [process (iii)→(i) in Fig. 1b]. The initial value of absorption intensity at 325 nm is completely recovered after about 10 minutes (Fig. 3d), related to the complete transformation of the molecules back to the original *EE*-form. This demonstrates both the expected reversibility of azobenzene photoisomerization in different environments,[46,47] and, more interestingly, the excellent photochemical properties of the axle which are retained in nanofibers. Fig. 3e shows the change in the intensity of the 325 nm peak upon five consecutive cycles of alternated irradiation periods with UV and blue laser light, evidencing fatigue resistance in photochemical cycling.

In electrospun nanofibers, both the polymer and embedded guest molecules can be stretched because of the applied electric field.[39,45,48-50] Polarized infrared spectroscopy can provide insightful information on the resulting molecular alignment. For instance, the Fourier-transform infrared (FTIR) absorption signal at 1730 cm$^{-1}$, arising from the C=O stretching of PMMA,[51] has a maximum for incident light polarized along the fiber axis, a result that evidences a partial alignment of the PMMA molecules (Fig. 4a). For this mode we measure a dichroic ratio (*R*)[39] of 1.5, calculated as $R = A_{\parallel}/A_{\perp}$, where $A_{\parallel}$ and $A_{\perp}$ are the absorption intensities for light polarized parallel and perpendicular to the fiber axis of alignment, respectively. A similar analysis is performed for the absorption peak at 1592 cm$^{-1}$, attributed to vibration of the C=C on the benzene ring[52] (Fig. 4a, inset) of the molecule **2** (see Fig. S2). The found dichroic ratio of ~ −0.73 for this signal is indicative of a partial alignment of the [*EE*-**1**H⊂**2**]•PF$_6$ complex along the fiber axis.

As the azobenzene photoisomerization could affect such a molecular arrangement, we analyzed the absorption intensity at 1592 cm$^{-1}$ varying the polarization angle of the incident light (Fig. 4b). The dependence of the transmitted intensity (*I*) of linearly polarized light upon passing through aligned fibers, on the angle (θ) formed by the directions of polarization of the incident light and the main axis of alignment, is roughly described by Malus' law, $I = I_0 \cos^2\theta + I_1$, where $I_1$ indicates the intensity of the unpolarized background. Following UV exposure, we do not find significant differences in the polarization curve. In fact, the relative positions of the azobenzene moieties with respect to the axle longitudinal direction are expected to change as a consequence of *E*→*Z* photoisomerization. Hence, our findings indicate that upon photoisomerization the transition moments related to the signal of **2** at 1592 cm$^{-1}$ are not significantly tilted with respect to the fiber longitudinal axis. This result suggests that the geometrical changes associated with isomerization have a negligible effect on the ring-axle supramolecular arrangement.

**Dethreading and rethreading by acid/base gas exposure**. In order to study the eventual dethreading and rethreading of the axle and ring in fibers, these are exposed to triethylamine (TEA) and HCl vapors for time intervals up to 4 hours, and the molecular assembly is extensively analyzed by PL spectroscopy. Upon UV excitation, the ring has a characteristic fluorescence emission peaked at about 310 nm,[53-55] whose intensity is completely quenched upon association with a molecular axle containing either *E*- or *Z*-azobenzene units (Fig. S3).

Conversely, the axle is not luminescent (Fig. S3). We firstly investigate the fluorescence properties of the pseudorotaxane system in solution, following base/acid treatment in order to induce the dethreading and rethreading of the ring and the axle.[41] The results of such an analysis (Fig. S4) show a quenching of the ring fluorescence upon formation of the pseudorotaxane complex, compared to a solution containing only the ring. The fluorescence intensity increases upon addition of TEA, suggesting that the deprotonated axle undergoes dethreading, and decreases again by adding HCl, in agreement with the rethreading of the protonated axle. The incomplete recovery of the initial emission intensity (Fig. S4) is most likely related to the fact that, in CHCl₃, the strongly coordinating $Cl^-$ anions compete with the ring for the ammonium site more efficiently than the $PF_6^-$ counterions.[56,57]

Therefore, the measurements of the ring PL provide a mean to investigate the dethreading and rethreading of the axle and ring. Here, this method is used to study such processes in solid state fibers, since the spectral features of the axle and the ring are preserved upon electrospinning the molecular components with PMMA (Fig. S5). Figure 5a shows the PL spectrum of the nanofibers embedding the [$EE$-$\mathbf{1}$H⊂$\mathbf{2}$]•$PF_6$ complex, which exhibits an emission band at $\lambda_{max}$ = 305 nm attributed to the uncomplexed ring molecules. The intensity of this emission band increases remarkably after exposure to TEA vapors (Fig. 5b) and decreases again to smaller values after subsequent exposure to HCl vapors (Fig. 5c). These results are consistent with a dethreading of the ring from the axle upon deprotonation of the ammonium site, induced by the TEA vapors [Fig. 1b, process (i)→(ii)], and with successive rethreading caused by acid treatment [Fig. 1b, process (ii)→(i)]. The observed increase of the ring PL intensity suggests that the TEA-induced dethreading is quantitative (Fig. 5a-b). In addition, the initial PL intensity is almost exactly restored upon HCl exposure (Fig. 5a-c), in contrast to what observed in solution (Fig.

S4a-c). Such a finding suggests that rethreading is more efficient in the fibers with respect to the organic solvent, possibly because the association of the chloride anions with the ammonium center, which hinders rethreading in organic solution, does not take place in the PMMA matrix.

In control experiments performed by exposing either (i) bare PMMA fibers or PMMA fibers embedding only the ring component to TEA, or (ii) PMMA/[$EE$-$\mathbf{1}$H$\subset$$\mathbf{2}$]•PF$_6$ fibers to N$_2$ (used as a carrier gas for TEA treatments), the samples do not show any variation of the intensity of the PL band attributed to the ring (Fig. S6). These results evidence that the change of the ring PL signal cannot arise from interactions of either the fiber matrix or the fluorescent component with the TEA molecules. Overall, our findings indicate that the exposure of [$EE$-$\mathbf{1}$H$\subset$$\mathbf{2}$]•PF$_6$ fibers to TEA vapors causes the disassembly of the threaded complex, which is reversibly re-assembled by treatment with HCl.

In order to have a more in-depth understanding of the dethreading/rethreading kinetics, [$EE$-$\mathbf{1}$H$\subset$$\mathbf{2}$]•PF$_6$-based fibers are exposed to TEA vapors and, successively, to HCl vapors for variable exposure times in a range comprised between 15 minutes and 4 hours. The results, shown in Fig. 5d-e, highlight that most of the PL intensity increase/decrease occurs within about 120 minutes. Such time-dependent changes can be attributed to the barrier associated with the dethreading/rethreading processes in the polymer matrix, combined with gas diffusion in electrospun fibers, which is known to occur on timescales of a few tens of minutes.[44] Furthermore, [$EE$-$\mathbf{1}$H$\subset$$\mathbf{2}$]•PF$_6$ fibers exposed firstly to UV laser light and then to TEA/HCl do not show significant differences (Fig. 5f-h) compared to samples directly treated with TEA/HCl (Fig. 5a-c), indicating that the dethreading/rethreading [processes (iii)→(iv) and (iv)→(iii) in Fig. 1b] of the ring and axle components in the fibers can occur even after the photoisomerization of the azobenzene units of the axle, as reported in solution.[41,42] Similarly, the $E{\rightarrow}Z$ photoisomerization of the azobenzene units of the axle in the fibers is not significantly affected by the TEA-induced dethreading, as shown in Fig. 6 wherein processes (ii)→(iv) and (iv)→(ii) of Fig. 1b are examined.

**Mechanical properties**. Finally, we investigate the possible corresponding variations of the *macroscopic* mechanical properties of the fibers due to the embedded pseudorotaxane components. The results reported in Fig. 7 clearly show that, while the mechanical properties of pristine PMMA nanofibers are in good agreement with previous reports,[58] the presence and – more interestingly – the dethreading of the axle and ring affect the mechanical properties of nanofibers. In particular, while the maximum strain (Fig. 7b) and ultimate tensile strength (Fig. 7c) show minor changes, large differences are found for the Young's modulus. The average Young's modulus of pure PMMA nanofibers aligned in arrays is here measured to be around $(50{\pm}10)$ MPa (not varying significantly following exposure to TEA), whereas it doubles in consequence of the embedment of either [*EE*-**1**H⊂**2**]•PF$_6$ (Fig. 7a) or its axle component alone (Fig. S7a). Upon TEA exposure, the Young's modulus of [*EE*-**1**H⊂**2**]•PF$_6$-doped fibers further increases up to $(160{\pm}20)$ MPa, indicating a significantly enhanced stiffness of the nanofibrous mat upon ring-axle dethreading. Instead, control experiments carried out on nanofibers with only the axle component do not lead to a similar increase of the Young's modulus (Fig. S7), thus ruling out a significant influence of the bare axle deprotonation on the mechanical properties at macroscale.

In addition, we find that the photoisomerization of the azobenzene end groups by exposure to UV light [namely, performing the (ii)→(iv) and (iv)→(ii) processes in Fig. 1b] does not affect significantly the fiber mechanical properties. Indeed, as shown in Fig. 8 the strain, the tensile strength and the modulus of [*EE*-**1**H⊂**2**]•PF$_6$-based fibers (both taken in their pristine state and following TEA treatment) remain almost unaltered after UV irradiation. An interpretation of these observations on the basis of the properties of the pseudorotaxane and its molecular components is not straightforward. It can be noticed, however, that the axle component, as well as the resulting ring-threaded species, are considerably more rigid than the PMMA molecules. It may therefore be argued that fibers in which these dopants are partially aligned with the host macromolecules could exhibit an increased stiffness, as observed in this work. Following TEA treatment and ring-axle dethreading, one also observes that the corresponding stress-strain curve becomes sub-linear (upward triangles in Fig. 7d), which suggests that the [*EE*-**1**H⊂**2**]•PF$_6$-doped fibers undergo microscopic rearrangements, and eventually plastic flow in their strain-hardening regime. This is especially interesting since the so obtained Young's modulus [(160±20) MPa] is significantly higher than the value found for fibers embedding the free axle [(100±10) MPa, Fig. S7]. Taken together, these results indicate that a specific increase of stiffness is associated to fibers embedding dethreaded molecular components. Issues related to the tensile properties of nanofibers and nanocomposites have been largely investigated for polymers embedding WS$_2$[58] or carbon[59-61] nanotubes, whereas the behavior of molecular blends and their nanostructures is rarely studied. In general, interfacial shear stresses between different components are at the base of load transfer to rigid fillers from the plastic matrix, thus determining improved mechanical properties in nanocomposites. Various mechanisms can improve load transfer, such as micro-mechanical interlocking between the host components and the fillers, chemical bonding, or van der Waals interactions,[59] and achieving a uniform dispersion of dopants in the polymer is also needed, as in our case. In particular, for the fibers studied in this work, dethreaded species can similarly promote load transfer from the plastic matrix due to van der Waals effects and correspondingly improved interfacial interactions between the polymer phase, or increasing the volume fraction of the dispersed component in the composite material, an effect which seems to take advantage of the presence of the separated axle *and* ring components. A full rationalization of the molecular details which lead to the observed mechanical properties will certainly deserve further investigations.

In general, the dependence of the mechanical properties of polymeric fibers on the switching of the supramolecular ring-axle interactions[62-64] is a significant step forward towards the exploitation of molecular-scale phenomena – including molecular movements – to bring about effects at the macroscopic level. This problem is indeed of the highest importance for the real world application of molecular devices and machines.[6-8] Materials like those described here are appealing for the development of sensing components working by chemical and photonic actuation.

### 3. SUMMARY

In summary, nanofibers embedding self-assembled rotaxane-type supramolecular complexes have been produced *via* electrospinning. These species are interesting because they are photoactive (azobenzene *E-Z* photoisomerization) and responsive to base-acid stimuli (chemically driven dethreading-rethreading). The possibility to reversibly interconvert the azobenzene units of the axle between their *E*- and *Z*-forms in the nanofibers has been assessed, as well as the reversible switching features brought about by dual (optical and chemical) stimulation of the embedded supramolecular system. Importantly, we demonstrated the influence of the ring and axle molecular components and of their dethreading on the macroscopic mechanical properties of fibers. These results are a step forward towards the incorporation of functional molecular machines into solid polymeric nanostructures, and their exploitation for the development of innovative devices and materials for photonics, (bio)chemical sensing, molecular release and mechanical actuation.

### 4. EXPERIMENTAL SECTION

**[EE-1H⊂2]•PF$_6$-based nanofibers.** Nanofibers are produced by electrospinning a 10% w/w
PMMA (120 kDa) solution in chloroform (Sigma Aldrich). The two pseudorotaxane components
are added at a total concentration of 42% w/w compared to PMMA (i.e., 15 mg of *EE*-**1**H•PF$_6$
and 48 mg of **2** in 150 mg of PMMA). Under these conditions the threaded complex [*EE*-**1**H⊂**2**]
•PF$_6$ is quantitatively afforded, as its stability constant in chloroform is of the order of $10^6$ M$^{-1}$.[41]

[*EE*-**1**H]•PF$_6$-based fibers are produced adding 15 mg of *EE*-**1**H•PF$_6$ in PMMA for control

mechanical experiments. The electrospinning apparatus comprises a syringe (1 mL, Hamilton)

and a 27 *gauge* stainless steel needle connected to a syringe pump (Harvard Apparatus) and a

high voltage supply (EL60R0.6-22, Glassman High Voltage Inc.). A positive voltage (5 kV) is

applied to the needle. The solution is supplied at a constant flow rate of 10 μL min$^{-1}$, and fibers

are collected on Al foils or on quartz cover slips. A copper plate biased at –6 kV and placed at 10

cm away from the needle is used as collector. Arrays of uniaxially aligned fibers are produced by

a metallic collector rotating at 2500 rpm. Electrospinning is performed at ambient conditions

(room temperature and relative humidity of about 50%). Fibers are stored at room temperature

before analysis. The morphology and the elemental composition of the pseudorotaxane-

embedding fibers are investigated by a scanning electron microscopy (SEM) system (Nova

NanoSEM 450, FEI) equipped with an energy dispersive X-ray spectrometer (Quantax, Bruker).

SEM images and elemental data are collected with an accelerating voltage of 1-5 kV and of 30 kV, and with an aperture size of 30 μm and 100 μm, respectively.

**Photoisomerization.** The $E{\rightarrow}Z$ photoisomerization of the azobenzene end groups is induced by exposure to UV light using the third harmonic of a pulsed Nd:YAG laser (Spectra Physics $\lambda$=355 nm, repetition rate=10 Hz, pulse duration=10 ns) for 0-20 min. The reversibility of the azobenzene photoisomerization (i.e. the $Z{\rightarrow}E$ conversion) is investigated by exposing samples to blue laser light (Micro Laser System, Inc.), with incident intensity 1.4 mW/cm$^2$ and $\lambda$=405 nm for different time intervals (0-10 min). Absorption spectra of the fibers in the spectral interval 250-400 nm are collected by a UV/visible spectrophotometer (Cary 300 Scan, Varian Inc.). Polarized FTIR spectroscopy is carried out using a spectrometer (Spectrum 100, PerkinElmer Inc.) equipped with an IR grid polarizer (Specac Limited, U.K.), consisting of 0.12 μm-wide Al strips. To this aim, freestanding samples of aligned fibers are analysed in transmission mode.

**Dethreading/rethreading experiments.** In order to induce the dethreading and rethreading of the ring and axle components, freestanding nanofibrous mats are exposed to a continuous nitrogen flow carrying TEA and HCl vapors, respectively. The used system is composed of two glass flasks connected through pipes. The first vial, containing 10 mL of TEA (or alternatively HCl), is connected to a controlled nitrogen flow (0.05 liters per minute), whereas the second bowl contains the sample. This is kept in the TEA (HCl) saturated nitrogen flux for variable exposure times in the range 15 minutes-4 hours, and investigated before and after TEA/HCl treatments by PL spectroscopy. To this aim the samples are excited by the fourth harmonic of the Nd:YAG pulsed laser ($\lambda$ = 266 nm) and the emission is collected by a quartz optical fiber, coupled to a monochromator (iHR320, Jobin Yvon) and a charged coupled device (CCD) detector (Symphony, Jobin Yvon). Since part of the incident light is diffused by the sample and collected by the optical fiber, a background spectrum is subtracted from each measured one (see Fig. S8). Dethreading and rethreading experiments are also performed in solution. To this aim, a solution containing *EE*-**1**H•PF$_6$ and **2** in 1.5 mL of chloroform is used, and dethreading and rethreading of the pseudorotaxane components are induced by adding 250 μL of TEA and HCl, respectively.

**Mechanical measurements.** Mechanical properties are determined using a dynamic mechanical analyzer (DMA Q800, TA Instruments, New Castle, DE). Each sample made of aligned nanofibers ($n = 5$ specimens) is cut into a rectangular shape (about 10×8 mm$^2$) before testing, and its thickness is measured using a digital micrometer (0.04-0.06 mm). Stress–strain curves are recorded with a ramp/rate of 1 N min$^{-1}$ (up to 18 N).

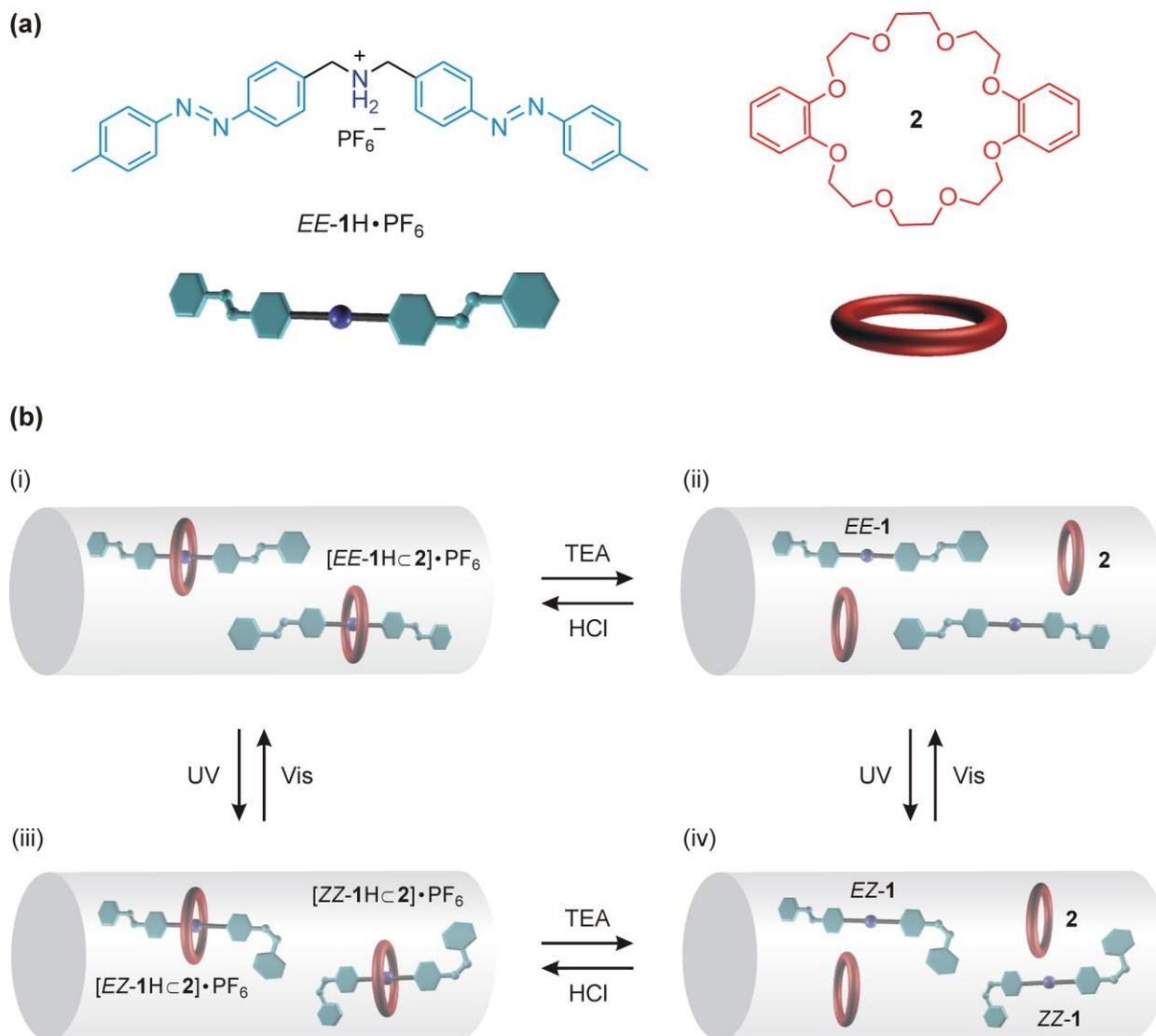

**Figure 1.** (a) Molecular structure of the axle (*EE*-1H•PF$_6$) and ring (**2**). (b) Schematic representation of the photochemical (*E-Z* isomerization of the azobenzene end units of the axle) and chemical (deprotonation/reprotonation-induced dethreading/rethreading) processes that the axle and ring molecular components can undergo inside the fibers.

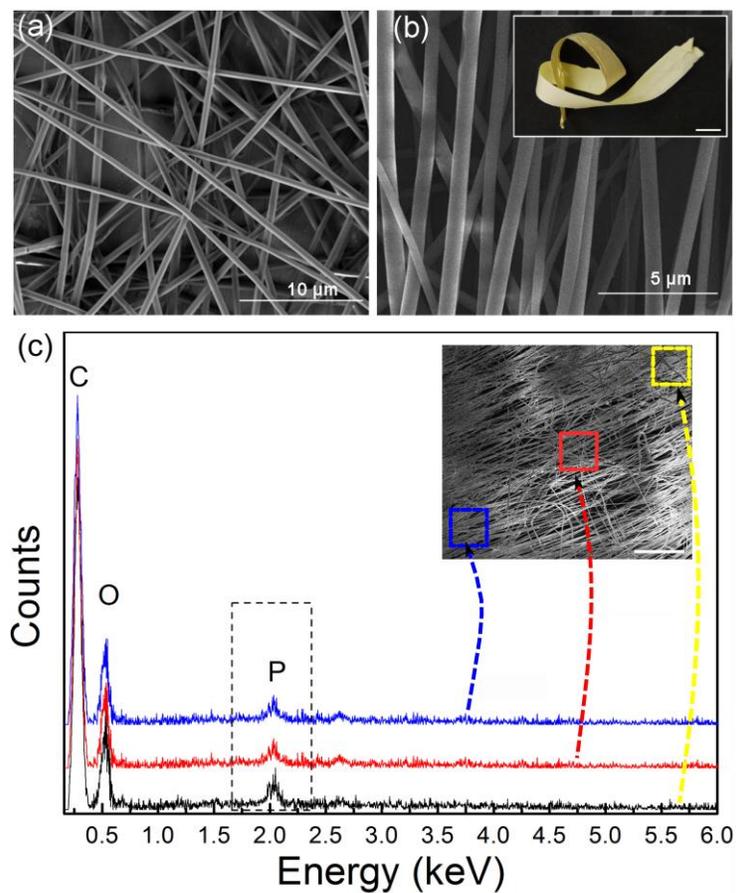

**Figure 2.** Scanning electron micrographs of randomly oriented (a) and prevalently aligned (b) PMMA/[*EE*-**1**H⊂**2**]•PF$_6$ fibers. In (b), about 70% of fibers have their longitudinal axis within 10° from the main alignment direction (see Supporting Information, Fig. S1). Inset: photograph of a uniaxially-oriented, [*EE*-**1**H⊂**2**]•PF$_6$-based nanofiber mat (scale bar: 1 cm). (c) EDS spectra of [*EE*-**1**H⊂2]•PF$_6$-based fibers measured in three different areas of the sample. The spectra are vertically shifted for better clarity. Inset: SEM micrograph of the fibers showing the 150×150 μm$^2$ investigated regions. Scale bar: 200 μm.

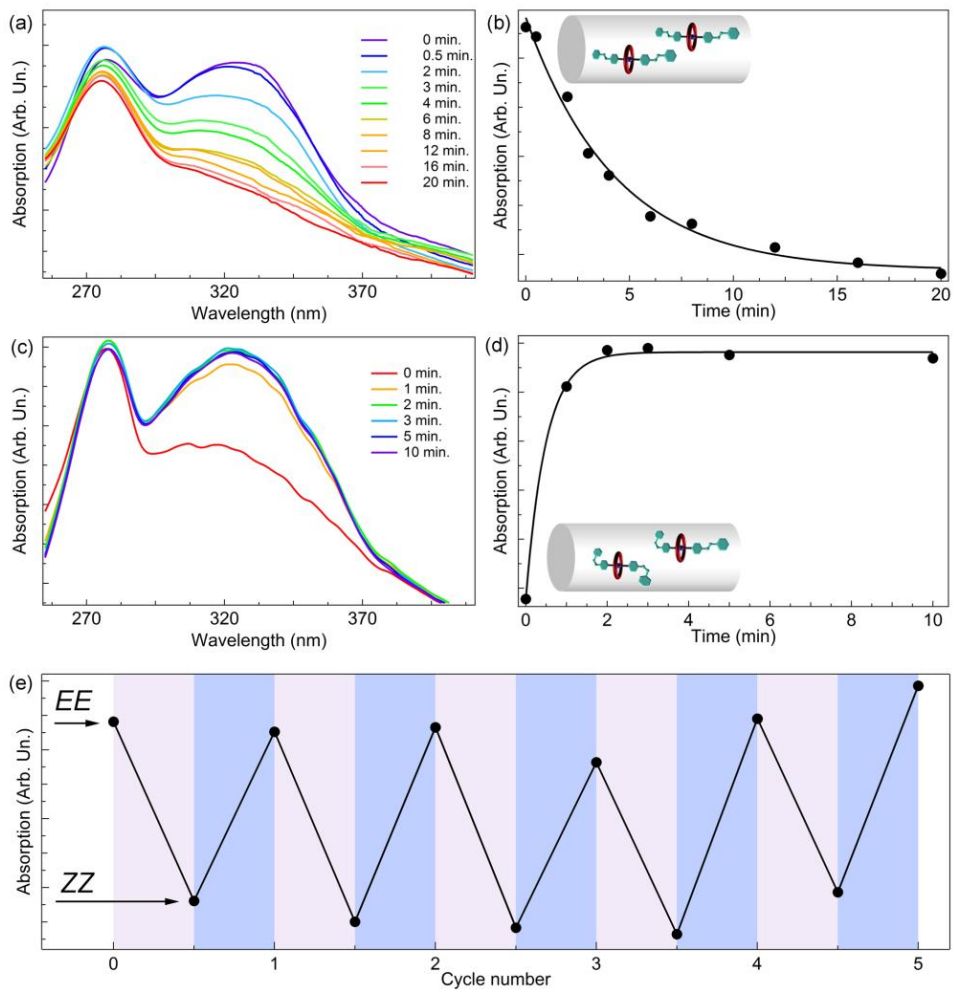

**Figure 3.** (a) Absorption spectra of [$EE$-**1**H⊂**2**]•PF$_6$-based nanofibers after different UV exposure intervals. Irradiation parameters: $\lambda$=355 nm, pulsed incident fluence = 60 µJ cm$^{-2}$. (b) Absorption values ($\lambda$ = 325 nm) vs. UV exposure time, corresponding to the spectra shown in (a). The continuous line is a guide for the eye. (c) Absorption spectra of fibers, previously subjected to exhaustive UV irradiation, for increasing time intervals of blue laser light exposure. Irradiation parameters: $\lambda$ = 405 nm, incident intensity = 1.4 mW cm$^{-2}$. (d) Plot of the absorption values ($\lambda$ = 325 nm) versus blue laser exposure time, corresponding to the spectra shown in (c). The continuous line is a guide for the eye. (e) Absorption changes at $\lambda$ = 325 nm, for consecutive cycles of alternated UV and blue laser irradiation.

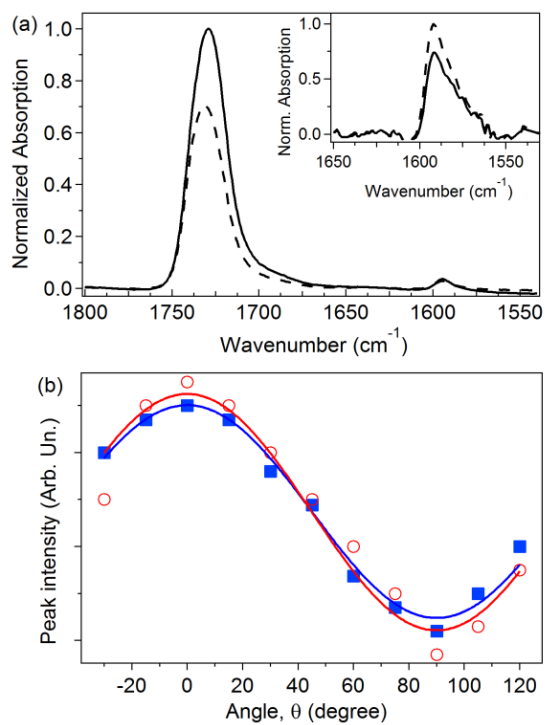

**Figure 4.** (a) FTIR spectra of [*EE*-**1**H⊂**2**]•PF$_6$-based fibers before UV exposure, taken at two angles of light polarization (0° and 90° with respect to fibers axis of alignment; continuous and dashed line, respectively). Inset: close-up of spectra around 1592 cm$^{-1}$. (b) Transmitted intensity of the peak at 1592 cm$^{-1}$ vs. angle (θ) formed by the polarization direction of incident light and the axis of alignment, before (full squares) and after (empty circles) UV exposure (pulsed incident fluence = 60 μJ cm$^{-2}$). Continuous lines represent the data fitting according to Malus' law.

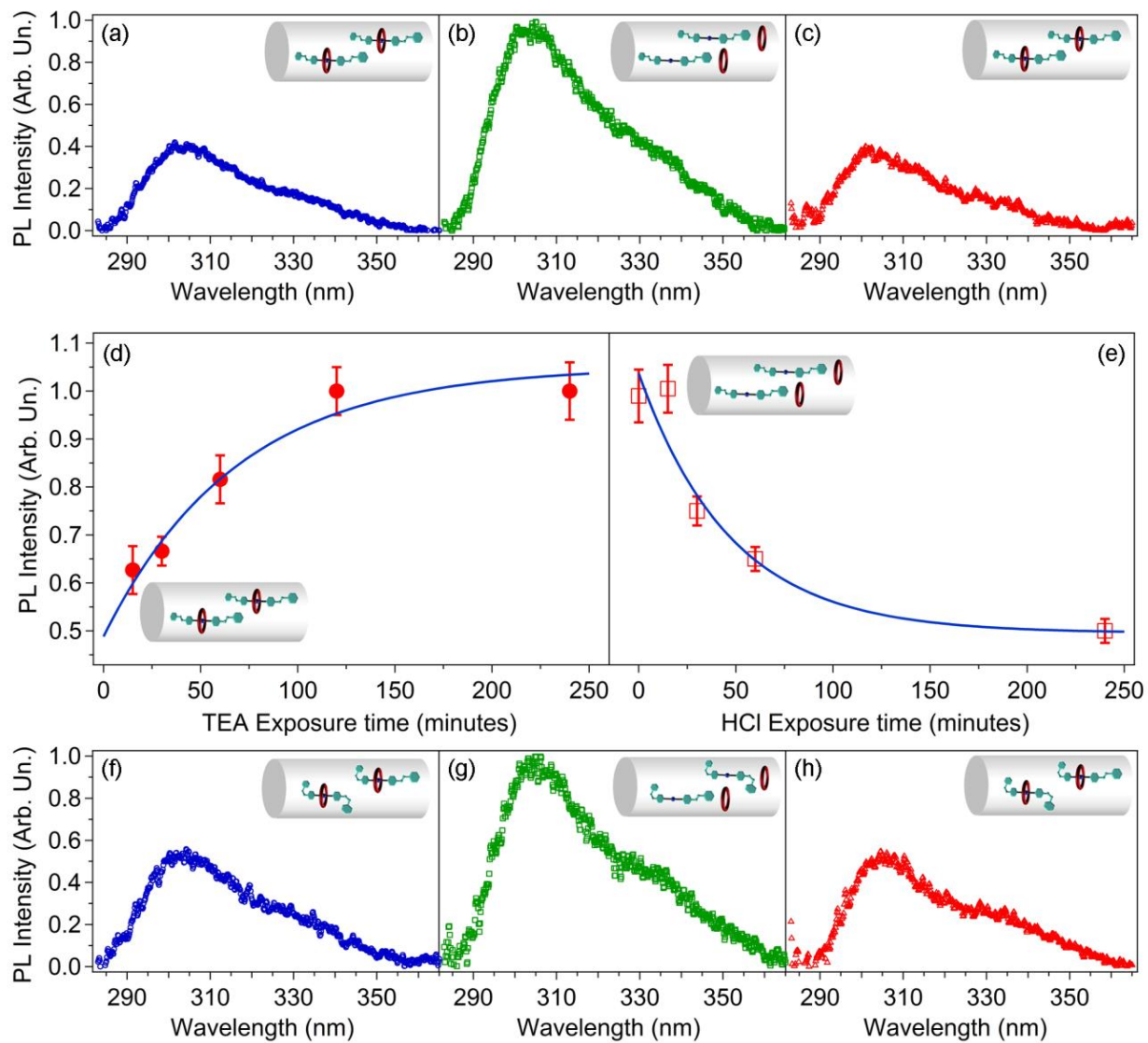

**Figure 5.** (a)-(c) PL spectra of [*EE*-**1**H⊂**2**]•PF$_6$-based fibers, before (a) and after (b) exposure to TEA vapors for 4 hours. (c) PL spectra of the samples shown in (a)-(b) after an additional exposure to HCl vapors for 4 hours. (d)-(e) PL intensity of [*EE*-**1**H⊂**2**]•PF$_6$-based fibers *versus* TEA (d) and HCl (e) exposure times. The samples used in (d) are firstly exposed to TEA vapors for 4 hours. Continuous lines are guides for the eye. (f)-(h) PL spectra of [*ZZ*-**1**H⊂**2**]•PF$_6$-based fibers PMMA, before (f) and after exposure to TEA (g) and HCl (h) vapors, respectively. In all cases, excitation was performed at 266 nm.

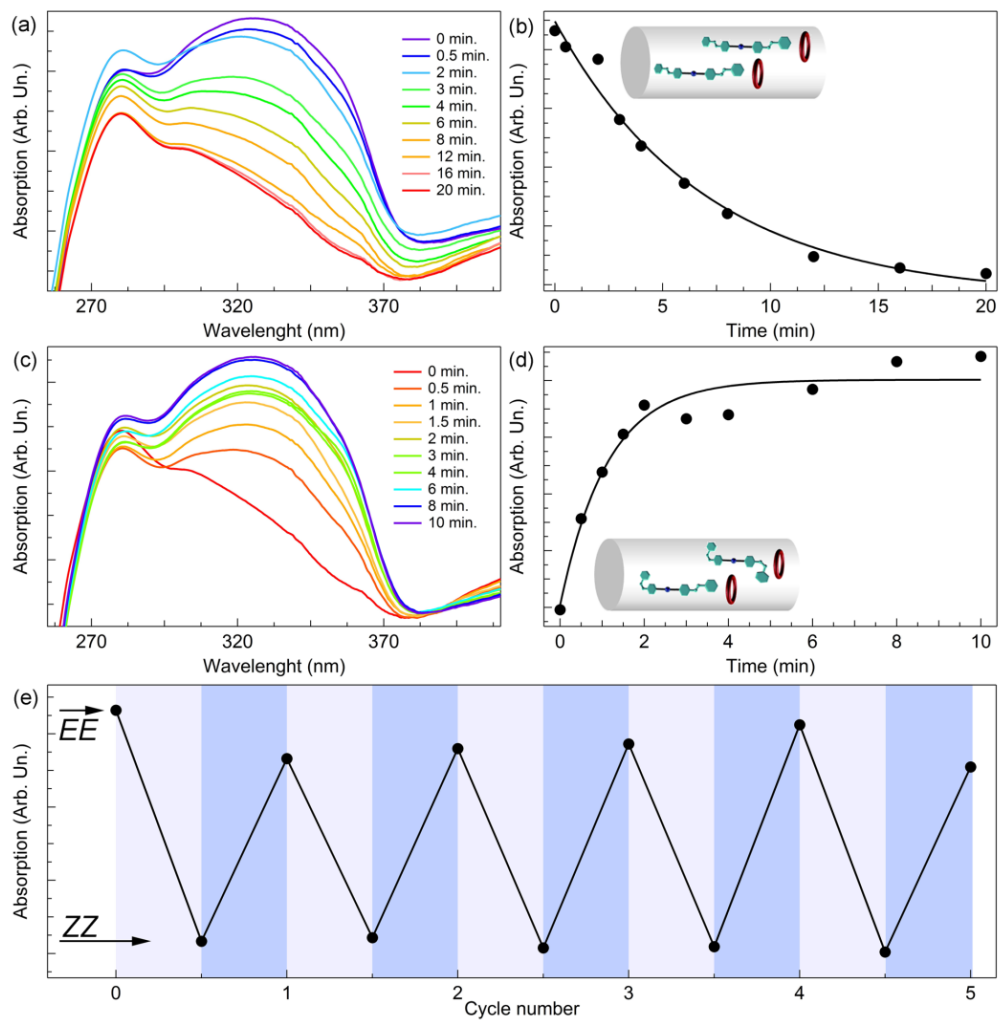

**Figure 6.** (a) Absorption spectra of [EE-**1**H⊂**2**]•PF$_6$-based nanofibers after exposure to TEA for 2 hours and subsequent, different UV exposure intervals. Irradiation parameters: $\lambda = 355$ nm, pulsed incident fluence = 60 μJ cm$^{-2}$. (b) Corresponding absorption values ($\lambda = 325$ nm) as a function of UV exposure times. The continuous line is a guide for the eye. (c) Absorption spectra of nanofibers, previously underwent UV irradiation, for increasing time intervals of blue laser light exposure. Irradiation parameters: $\lambda = 405$ nm, incident fluence = 1.4 mW cm$^{-2}$. (d) Corresponding absorption values ($\lambda = 325$ nm) as function of exposure time. The continuous line is a guide for the eye. (e) Plot of the absorption changes at $\lambda = 325$ nm, for consecutive cycles of alternated UV and blue laser irradiation.

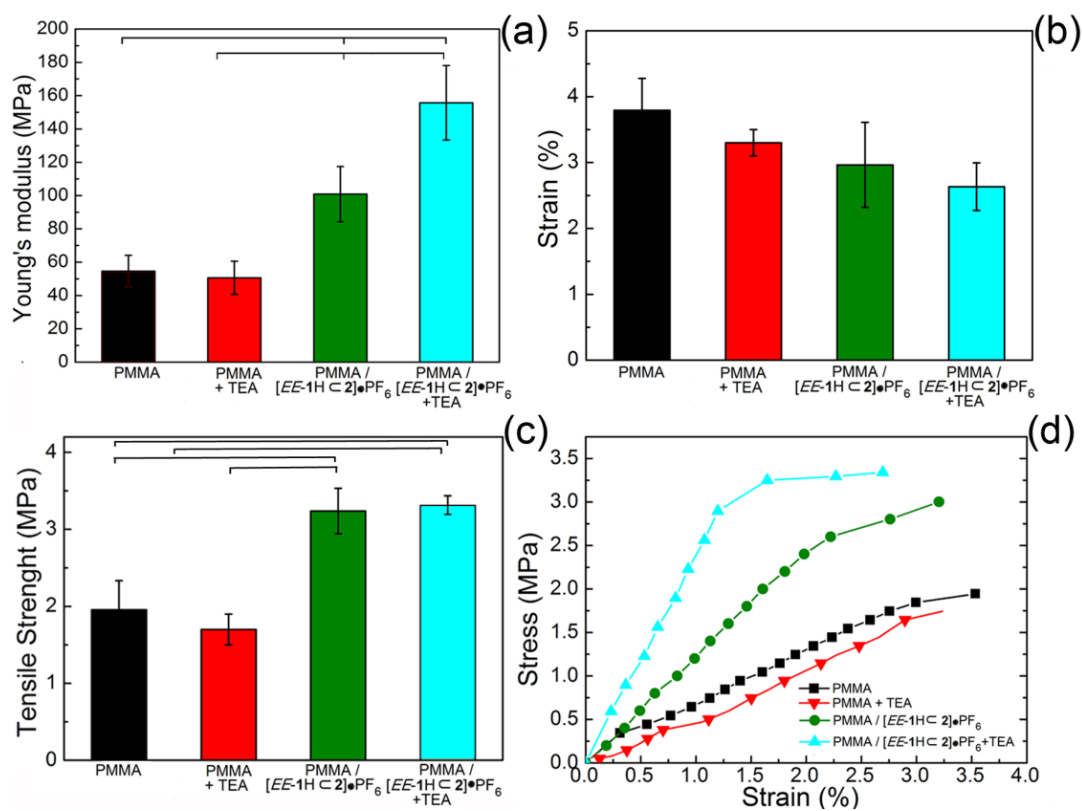

**Figure 7.** Mechanical properties of PMMA and PMMA/[*EE*-**1**H⊂**2**]•PF$_6$ nanofibers before and after TEA exposure. (a) Young's Modulus; (b) Strain and (c) Tensile Strength. (d) Representative Stress/Strain curves. Results are expressed as mean ± standard deviation. Bars show statistically significant differences (*P*<0.05).

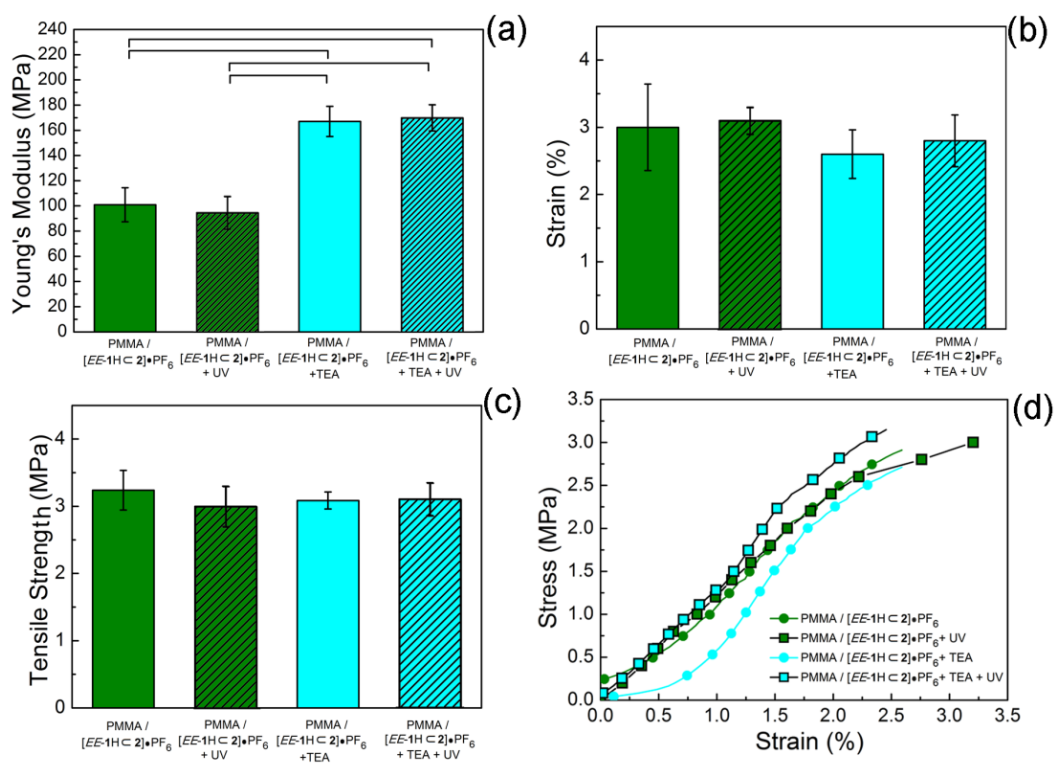

**Figure 8**. Mechanical properties of PMMA/[*EE*-**1**H⊂**2**]•PF$_6$ and PMMA/[*EE*-**1**H⊂**2**]•PF$_6$ exposed to TEA vapors for 2 hours, before and after UV exposure. (a) Young's Modulus; (b) Strain and (c) Tensile Strength. (d) Representative Stress/Strain curves. Results are expressed as mean ± standard deviation. Bars show statistically significant differences (*P*<0.05).

ASSOCIATED CONTENT

**Supporting Information**. A supporting document with additional technical details on nanofiber morphological properties, and on spectral and mechanical characterization is included as a separate PDF file. This material is available free of charge via the Internet at http://pubs.acs.org.

AUTHOR INFORMATION

**Corresponding Authors**

* E-mail: dario.pisignano@unisalento.it, alberto.credi@unibo.it, andrea.camposeo@nano.cnr.it

**Present Addresses**

§ Current address: Laboratory of Organic Electronics, Department of Science and Technology

Linköping University, Norrköping, 601 74 Sweden.

ACKNOWLEDGMENT

We thank Dr. Serena Silvi for useful discussions. We acknowledge the support from the Italian

Ministry of University and Research (FIRB contract RBNE08BNL7-MERIT Program, and PRIN

2010CX2TLM), from the University of Bologna (FARB project "SLaMM"), and from the

Apulia Regional Projects 'Networks of Public Research Laboratories' Wafitech (9) and M. I. T.

T. (13). The research leading to these results has received funding from the European Research

Council under the European Union's Seventh Framework Programme (FP/2007-2013)/ERC

Grant Agreement n. 306357 (ERC Starting Grant "NANO-JETS").

# SUPPORTING INFORMATION

# Organic nanofibers embedding stimuli-responsive threaded molecular components

Vito Fasano, [†] Massimo Baroncini, [‖] Maria Moffa, [#] Donata Iandolo, [#] Andrea Camposeo,[#,*] Alberto Credi [‖,*] and Dario Pisignano[†,#,*]

[†] Dipartimento di Matematica e Fisica "Ennio De Giorgi", Università del Salento, via Arnesano I-73100 Lecce (Italy)

[#] National Nanotechnology Laboratory of Istituto Nanoscienze-CNR, via Arnesano, I-73100 Lecce (Italy)

[‖] Photochemical Nanosciences Laboratory, Dipartimento di Chimica "G. Ciamician", Università di Bologna, via Selmi 2, I-40126 Bologna (Italy)

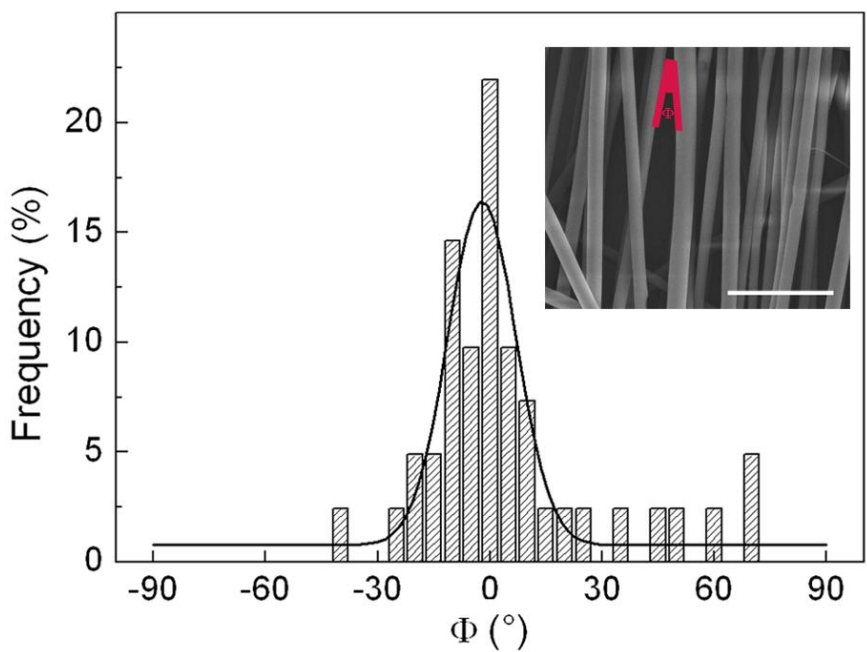

**Figure S1.** Angular distribution of aligned PMMA/[*EE*-**1**H⊂**2**]•PF$_6$ nanofibers. The alignment degree is evaluated measuring the angle formed by each fiber with respect to the prevalent alignment direction. The continuous line is the Gaussian fit of data. Inset: SEM image of aligned fibers. Φ: angle formed by each fiber with respect to an arbitrary vertical axis. Scale bar: 5 μm.

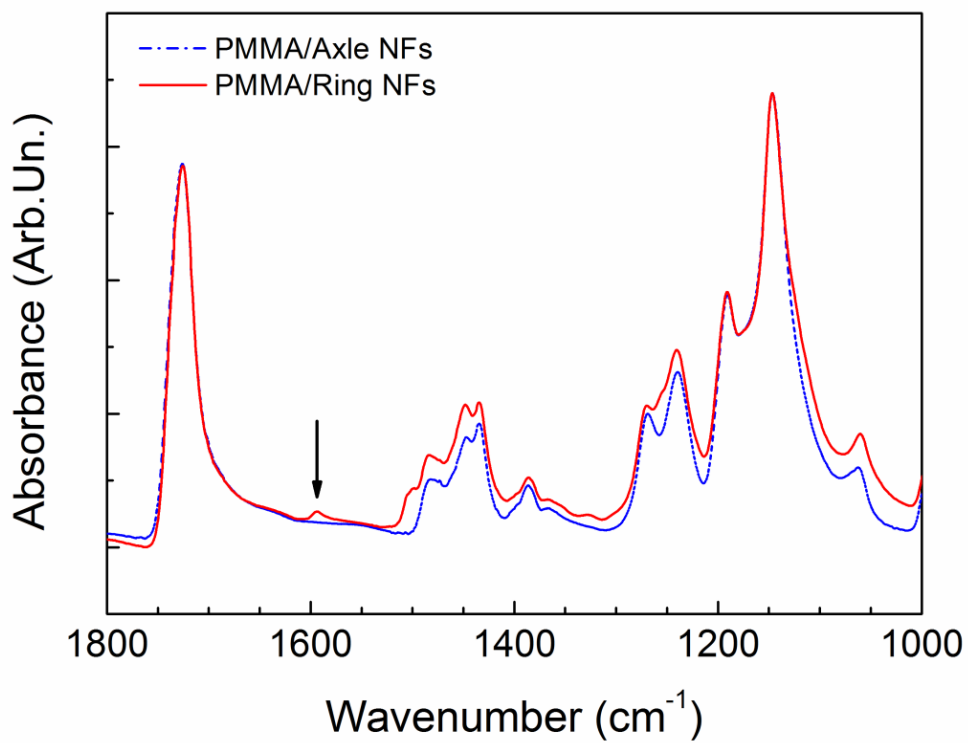

**Figure S2.** Horizontal Attenuated Total Reflection-FTIR spectra of PMMA nanofibers (NFs) embedding the axle (*EE*-**1**H•PF$_6$, dash-dotted blue line) and the ring (**2**, continuous red line). The vertical arrow highlights the mode at about 1592 cm$^{-1}$.

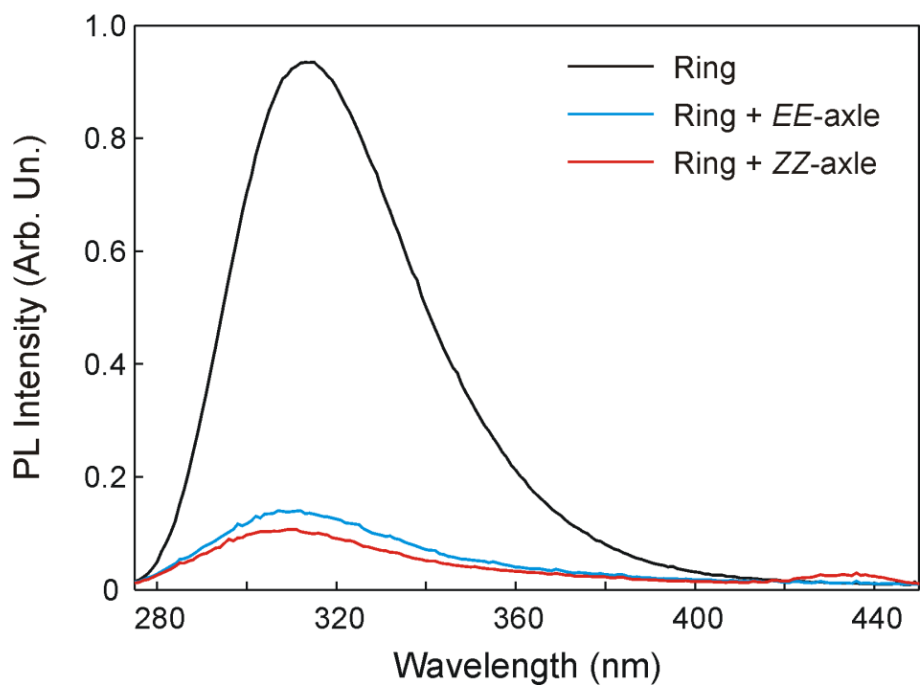

**Figure S3.** Photoluminescence spectra ($\lambda_{exc}$ = 260 nm) of an air equilibrated $CH_2Cl_2$ solution containing the ring **2** alone (20 μM, black line) and in the presence of either *EE*-**1**H•$PF_6$ (blue line) or *ZZ*-**1**H•$PF_6$ (red line). The *ZZ* isomer of the axle was obtained *in situ* by exhaustive irradiation at 365 nm of the corresponding *EE* form (see Ref. 40 in the main text). The spectra are corrected for inner filter effects, as the azobenzene units of the axle reabsorb part of the light emitted by the ring. The residual luminescence intensity in the presence of the axles (blue and red curves) is ascribed to uncomplexed **2**, in agreement with an association constant between the ring and the secondary ammonium axles in the order of $10^6$ $M^{-1}$ (Refs. 52, 54 in the main text). Hence, these experiments indicate that the luminescence of the ring is completely quenched when it is complexed by either axle. Compounds *EE*-**1**H•$PF_6$ and *ZZ*-**1**H•$PF_6$ are not luminescent under the same experimental conditions.

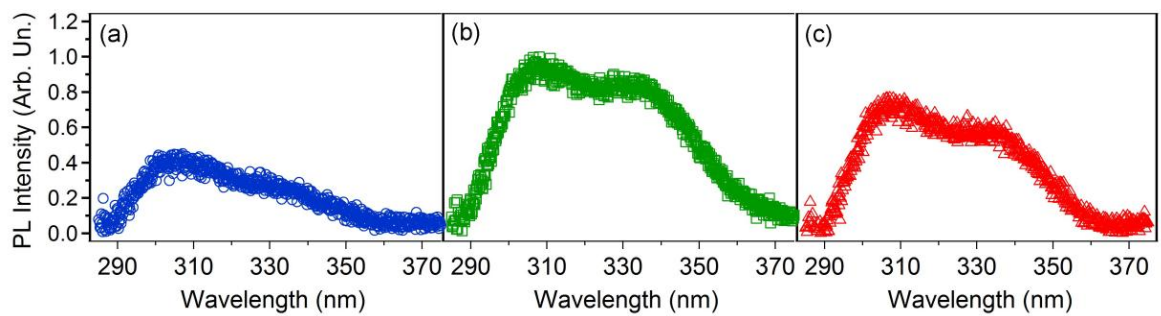

**Figure S4.** Photoluminescence spectra of a chloroform solution containing *EE*-**1**H•PF$_6$ and **2** (a), and of the same solution after the addition of TEA (b) and HCl (c), respectively.

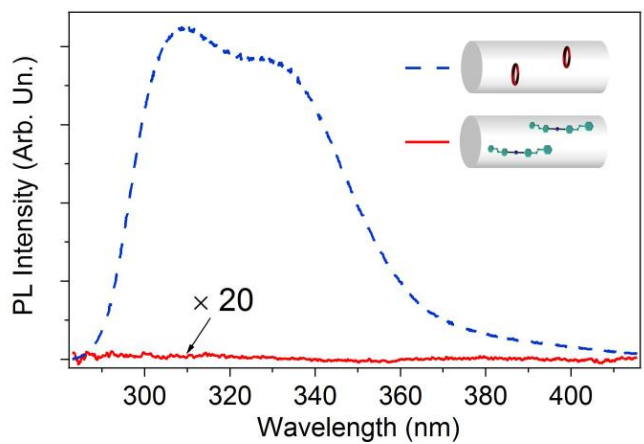

**Figure S5.** Photoluminescence spectra of PMMA electrospun nanofibers containing the axle, *EE*-**1**H•PF$_6$ (red continuous line) and the dibenzo[24]crown-8 molecular ring **2** (blue dashed line). Excitation was performed at 266 nm.

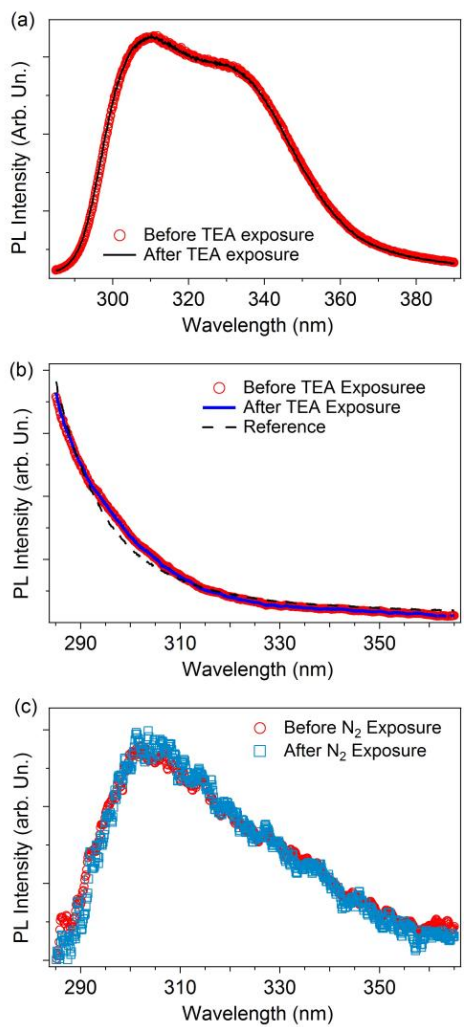

**Figure S6.** (a) Photoluminescence spectra of PMMA nanofibers embedding the dibenzo[24]crown-8

molecular ring before (red circles) and after (black continuous line) exposure to TEA vapors. (b)

Spectra of PMMA nanofibers before (red circles) and after (black continuous line) exposure to TEA

vapors, showing no significant differences. A reference signal collected from a quartz window is also

shown as reference (dashed line), corresponding to the background originating from diffusion of the

excitation laser light. The spectra collected from fibers do not significantly differ from the reference

one, ruling out the presence of photoluminescence contributions coming from the nanofiber polymer

matrix. (c) Photoluminescence spectra of PMMA nanofibers containing the pseudorotaxane [*EE-*

**1**H⊂**2**]•PF$_6$ complex, before (red circles) and after (blue squares) exposure to N$_2$. In all cases, excitation was performed at 266 nm.

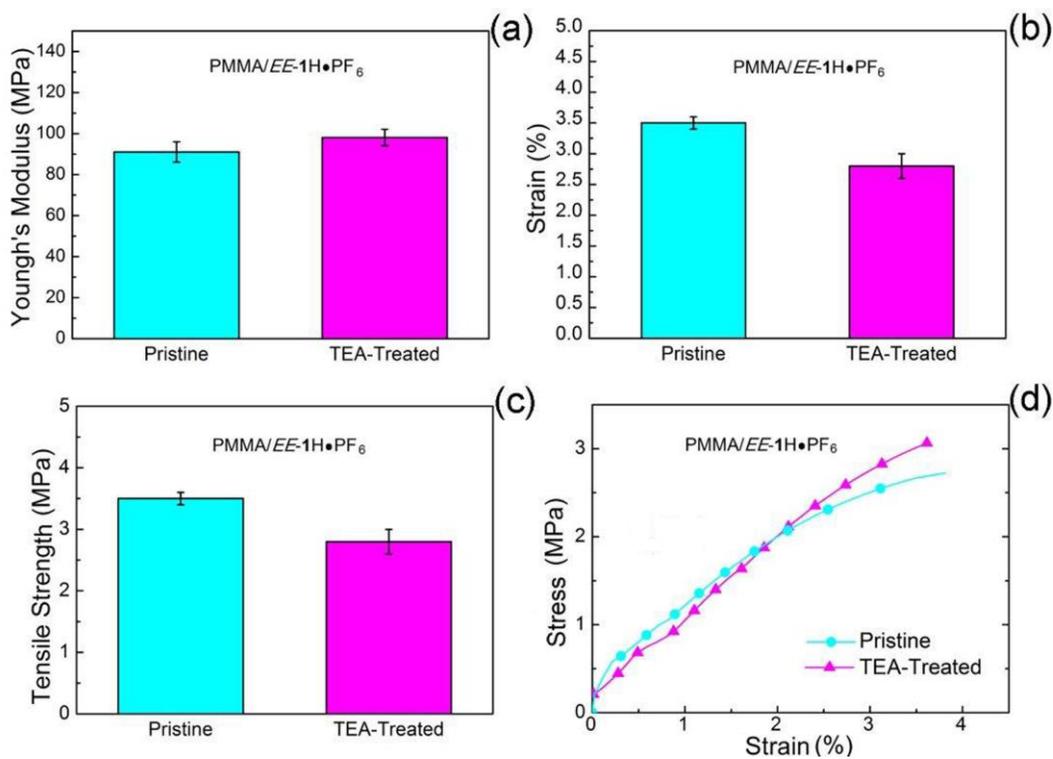

**Figure S7.** Mechanical properties of *EE*-**1**H•PF$_6$-based nanofibers before and after TEA exposure. (a) Young's Modulus; (b) Strain and (c) Tensile Strength. (d) Representative Stress/Strain curves. Results are expressed as mean ± standard deviation.

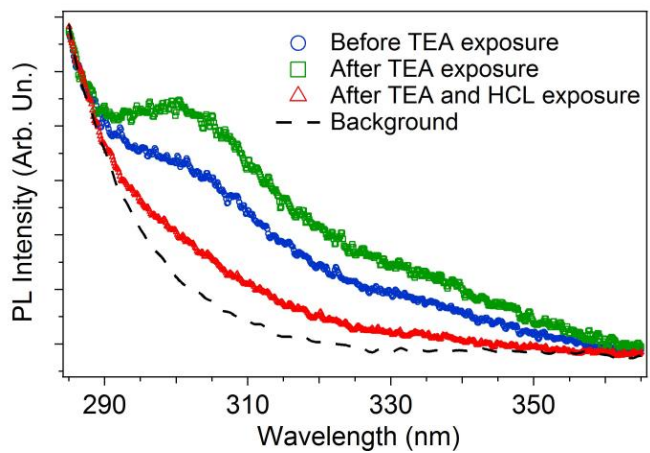

**Figure S8.** Representative experimental photoluminescence spectra acquired on PMMA nanofibers embedding the pseudorotaxane [*EE*-**1**H⊂**2**]•PF$_6$ ($\lambda_{exc}$ = 266 nm). Circles: pristine fibers; squares: after TEA treatment; triangles: after consecutive TEA and HCl treatment. A reference signal collected from a quartz window is also shown as reference (dashed line), highlighting the background signal due to diffusion of the excitation laser light. The photoluminescence spectra reported in the paper are obtained by subtracting the background contribution from each acquired spectrum.